\documentclass[12pt,a4paper]{article}
\pdfoutput=1
\usepackage{amsmath,amssymb,exscale}
\usepackage{amsfonts,amsthm}
\usepackage{graphicx, float, epsfig}
\usepackage[export]{adjustbox}
\usepackage{mathrsfs}
\usepackage{slashed}
\usepackage{textcomp}
\usepackage[T1]{fontenc}
\usepackage[utf8]{inputenc}
\usepackage[table]{xcolor}
\usepackage[lowtilde]{url}
\usepackage{cancel}
\usepackage[colorlinks=true,urlcolor=blue,anchorcolor=blue,citecolor=blue,filecolor=blue,linkcolor=blue,menucolor=blue,pagecolor=blue,linktocpage=true]{hyperref}
\numberwithin{equation}{section}
\usepackage{verbatim} 
\usepackage{appendix}
\newcommand{\be}{\begin{equation}}
\newcommand{\ee}{\end{equation}}
\newcommand{\ba}{\begin{eqnarray}}
\newcommand{\ea}{\end{eqnarray}}
\newcommand{\bay}{\begin{array}{rcl}}
\newcommand{\eay}{\end{array}}

\begin{document}

\title{A Stringy Model of Pointlike Particles}

\author{Risto Raitio\footnote{E-mail: risto.raitio@gmail.com}\\	
02230 Espoo, Finland}

\date{May 12, 2021}  \maketitle 

\abstract{\noindent  
A previous supersymmetric preon scenario for visible matter particles is extended to the dark sector. In addition, the scenario is reformulated as a Double Field Theory (DFT) with four extra dimensions, to avoid a singular Big Bang in cosmology. T-duality and doubled local Lorentz symmetry of the model are genuine stringy properties. It is proposed that DFT preons may be an approximate pointlike projection of string theory.}



\vskip 0.5cm
\noindent
Published in 	Nuclear Physics B980 (2022) 115826  \\
DOI				\href{https://doi.org/10.1016/j.nuclphysb.2022.115826} {\tt10.1016/j.nuclphysb.2022.115826}				
\newpage

\tableofcontents
\vskip 1.0cm

\section{Introduction}
\label{intro}

A previous supersymmetric preon scenario is (i) extended to include the dark sector, and (ii) reformulated as a Double Field Theory (DFT) in four plus four dimensions. The role of the preons is that they open a door to go beyond the standard model up to supergravity. A connection to bosonic strings is anticipated. 

Preons are free particles above an energy scale $\Lambda_{cr}$. This scale is estimated to be close (but above) to what is usually called the reheating temperature in the early universe, about $T_R \sim 10^{10-13}$ GeV. Below $\Lambda_{cr}$ preons make a phase transition into pointlike composite states of quarks and leptons at current accelerator energies and far beyond. 
 
Visible matter can be built from two preons (of charges $\frac{1}{3}$ and 0) and their antiparticles. Preons have only gravitational and electromagnetic interactions.\footnote{Gravity and electromagnetism with two fermions can make a perfect universe. With the SM non-Abelian sector chemistry and biology become available providing the universe with observers. Above $\Lambda_{cr}$ the non-Abelian couplings are small anyhow.} Their baryon (B) and lepton (L) numbers are zero. The dark sector consists of one zero charge fermion and axion like particles (ALP).

The extension to DFT is required by cosmological observations, like avoiding a singular Big Bang. In addition, to approach a unified theory a possible connection of preons to string theory is disclosed. 

This short note is organized as follows. In section \ref{darkmatter} we recap, refine and extend our previous preon model. In section \ref{doublefieldtheory} we introduce the new symmetries to the model, like Double Field Theory in four plus four dimensions, generalized diffeomorphism and double local Lorentz symmetry. The symmetries of the extended model are discussed in section  \ref{extsymmod}. Conclusions are given in section \ref{conclusions}.

\section{Supersymmetric Visible and Dark Matter}
\label{darkmatter}

The starting point in the preon model for visible and dark matter is that supersymmetry must be defined so that superpartners are included in the model on the first line together with the base particles. Put it differently, if one removes supersymmetry from the model either fermions or bosons are lost from the model. This is the case in superstring theory: if supersymmetry is removed one gets only bosonic strings. In the Minimal Supersymmetric Standard Model (MSSM), on the contrary, if one removes  supersymmetry from it one gets the SM with all the known fermions and bosons observed in laboratory experiments. 

{\bf Visible Matter.} Compared to the SM, this preon model is in its infancy.\footnote{Preons were studied actively in the 1980's, even sypersymmetric ones, but our model is different.} Instead of the internal symmetries our starting point is global supersymmetry with mass and charge.\footnote{The usual internal symmetries are considered emergent.} We briefly recap the phenomenological preon scenario of \cite{Rai_00, Rai_01}, which turned out to have close resemblance to the simplest N=1 globally supersymmetric 4D model, namely the free, massless Wess-Zumino model \cite{Wess_Z, Figueroa-O} with the kinetic Lagrangian including three neutral fields $m$, $s$, and $p$ with $J^P = \frac{1}{2}^+, 0^+$, and $0^-$, respectively
\be
\mathcal{L}_{\rm{WZ}} = -\frac{1}{2} \bar{m}\cancel\partial m - \frac{1}{2} (\partial s)^2 - \frac{1}{2} (\partial p)^2
\label{chirallagrangian}
\ee
where $m$ is a Majorana spinor, $s$ and $p$ are real fields (metric is mostly plus). 

We assume that the pseudoscalar $p$ is the axion \cite{Peccei_Q}, and denote it below as $a$. It has a fermionic superparther, the axino $n$, and a bosonic superpartner, the saxion $s^0$.

In order to have visible matter we assume the following charged chiral field Lagrangian
\be
\mathcal{L}_{-} = -\frac{1}{2}m^- \cancel\partial m^- -\frac{1}{2}(\partial s^-_i)^2, ~~i=1,2
\label{chargedchiral}
\ee
The R-parity of preons is simply $P_R = (-1)^{2\times spin}$.

The first generation standard model quarks and leptons are introduced as composite states of three preons: the charged $m^{\pm}$, with charge $\pm\frac{1}{3}$, and the neutral $m^0$. The compositeness may be due to trefoil states \cite{Finkelstein}, or (unspecified) gauge theory bound states. The composite states, which occur below an energy scale $\Lambda_{cr}$, are listed in the upper part of Table \ref{tab:table0}. The table indicates that permutations are defined for preons to provide color. Unlike charged preon pairs, $m^+m^-$ are not included as they would annihilate into radiation and prevent forming a three preon composite state. The preon mutual interactions are at this stage unknown.  Preon confinement looks like quark confinement though the spectrum is not known. 

The composite states of preons are below the energy scale $\Lambda_{cr}$ to a good approximation pointlike, say close to the electron Cartan radius. The gluons and weak gauge bosons see the composite quarks and gluons as pointlike particles. 

{\bf Dark Matter.} Literature on dark matter, dark energy, and axions is extensive, see e.g. \cite{Marsh_00, Marsh_01, Baer_C_K_R, Marsh_et_al, DiBari, Cline}. In this section we patch our shortage in \cite{Rai_01} to consider the pseudoscalar of (\ref{chirallagrangian}). So we start from the Lagrangian (\ref{chirallagrangian}).

The superpartners of the axion $a$ are the fermionic axino $n$, and the scalar saxion $s^0$, also indicated in Table \ref{tab:table0}.\footnote{In this note we mostly talk about all spin zero particles freely as scalars.} Particle dark matter consists of all these three particles. The axino $n$ may appear physically as single particle dust or three $n$ composite $o$ dust, gas, or a large astronomical object. The fermionic DM behaves naturally very differently from bosonic DM, which may form in addition Bose-Einstein condensates.

Other candidate forms of DM include primordial black holes (PBH). They can be produced by gravitational instabilities induced from scalar fields such as axion-like particles or multi-field inflation. It is shown in \cite{Ketov_A_A} that PBH DM can be produced only in two limited ranges of $10^{-15}$ or $10^{-12}$ Solar masses ($2 \times 10^{30}$ kg). Dark photons open a rich phenomenology described \cite{Fayet}. 

\begin{table}[h!]
  \begin{center}
    \begin{tabular}{|l|l|} 
	\hline
     SM Matter & Preon state \\ 
	\hline
      $\nu_e$ & $m^0 m^0 m^0$ \\
	 $u_R$ & $m^+ m^+ m^0$ \\
	 $u_G$ & $m^+ m^+ m^0$ \\  
	 $u_B$ & $m^+ m^+ m^0$ \\

	 $e^-$ & $m^- m^- m^-$ \\
	 $d_R$ & $m^- m^0 m^0$ \\
	 $d_G$ & $m^- m^0 m^0$ \\
	 $d_B$ & $m^- m^0 m^0$ \\
	 \hline
     Dark Matter & Preon state  \\
	 \hline
	 boson (or BC) & axion(s), $s^0$ \\
	 $e'$ & axino $n$ \\
	 meson, baryon $o$ & $n\bar{n}, 3n$ \\
	 nuclei (atoms with $\gamma ')$ & multi $n$ \\
	 celestial bodies & any dark stuff \\	 
	 black hole & any preon \\
	\hline
    \end{tabular}
    \caption{\footnotesize{Visible and Dark Matter with corresponding particles. Color is carried only by the neutral $m^0$ preon. The visible matter preon states do not allow (annihilating) unlike charge $m^+m^-$ pairs. BC stands for Bose condensate. $e'$ and $\gamma '$ refer to dark electron and dark photon, respectively.}}
    \label{tab:table0}
  \end{center}
\end{table}

The axion was originally introduced to solve the strong CP problem in quantum chromodynamics (QCD) \cite{Peccei_Q}, see also \cite{Wilczek, Weinberg_ax}. The PQ axion has a mass in the range $10^{-5} ~ \rm{eV~ to}~ 10^{-3}$ eV. Axions, or axion-like particles (ALP), occur also in string theory in large numbers (in the hundreds), they form the axiverse. 

The axion-like particle masses extend over many orders of magnitude making them distinct candidate components of dark matter. Ultra-light axions (ULA), with masses $10^{-33}$ eV $< M_a < 10^{-20}$ eV, roll slowly during inflation and behave like dark energy before beginning to oscillate (as we see below). The lightest ULAs with $M_a \lessapprox 10^{-32}$ eV are indistinguishable from dark energy. Higher mass ALPs, $M_a \gtrapprox 10^{-25}$ eV behave like cold dark matter \cite{Marsh_et_al}. Quantum mechanically, an axion of mass of, say $10^{-22}$ eV, has a Compton wavelength of $10^{16}$ m.

Ultra-light bosons with masses $\ll$ eV can form macroscopic systems like Bose-Einstein condensates, such as axion stars \cite{Guerra_M_P, Annulli_C_V_0}. Due to the small mass the occupation numbers of these objects are large, and consequently, they can be described classically. 

The fermionic axino $n$ is supposed to appear, like the $m$ preons, as free particle if $T > \Lambda_{cr}$ and when $T \lesssim \Lambda_{cr}$ in composite states. If the mass of the three axino composite state $o$ is closer to the electron mass, or more, rather than the neutrino mass it may form 'lifeless' dark stars in a wide mass range. There are several possibilities for dark matter constituents to form bound states both with Abelian  and non-Abelian gauge groups leading to dark atoms, mesons and baryons, for a review see \cite{Cline}. 

As an estimate of the strength of the DM self-interaction we cite the upper limit on it \cite{Cline, Steinhardt_S, Steinhardt_W_D_F_M_S}
\be
\frac{\sigma}{M} \sim 0.5 \frac{\rm{cm^2}}{\rm{g}} 
\label{selfinteractionstrength}
\ee
Saturating this limit helps to ameliorate small scale problems like the cusp-core issue. 

The cross section (\ref{selfinteractionstrength}) is very large for an elementary DM particle of mass $M \sim 1$ GeV, but it is easy to achieve with composite particles. For dark atoms we can expect a geometric cross section governed by the dark Bohr radius, $a'_0 = 1/(\alpha' \mu_{H'})$, where $\alpha'$ is the dark coupling strength, $\mu_{H'}$ the reduced dark mass and $\sigma \sim \pi a'^2_0$.

\section{Double Field Theory}
\label{doublefieldtheory}

The mathematical formalism in this section follows closely the Ringberg Castle lectures \cite{Hassler_L} and thesis \cite{Hassler_T} of Hassler.

\subsection{Action}
\label{action}

Consider field theory in N (below N = 4) dimensional compact spacetime. Double Field Theory in 2N dimensions can take into account both momentum, and another quantity, the winding modes of a closed loop. Therefore DFT is a low energy effective theory of closed string theory. In addition of the spacetime coordinates $x^i$, which are conjugate to the momentum modes, there are now N new coordinates $\tilde{x}_i$, which are in turn conjugate to the winding modes of the closed string. We now deal with 2N dimensional vectors $X^M = (\tilde{x}_i, x^i)$. The index M is lowered and raised by the O(D,D) invariant matrix 
\be
\eta_{MN} = \left( \begin{array}{cc} 0 & \delta_i^j \\
\delta_j^i & 0 \end{array} \right) \rm{and~ its~ inverse} ~~
\eta^{MN} = \left( \begin{array}{cc} 0 & \delta_j^i \\
\delta_i^j & 0 \end{array} \right) 
\label{oddinvmatrix}
\ee

The action of a DFT in the generalized matrix formulation is \cite{Hohm_H_Z}
\be
S_{DFT} = \int \rm{d}^{2D} e^{-2d}\mathcal{R}
\label{dftaction}
\ee
where
\begin{align}\label{genricciscalar}
  \mathcal{R} = 4 \mathcal{H}^{MN} \partial_M d \partial_N d
    - \partial_M \partial_N \mathcal{H}^{MN} &- 4\mathcal{H}^{MN}
      \partial_M d \partial_N d + 4 \partial_M \mathcal{H}^{MN}
      \partial_N d \nonumber \\
  + \frac{1}{8} \mathcal{H}^{MN} \partial_M \mathcal{H}^{KL}  
    \partial_N \mathcal{H}_{KL} &- \frac{1}{2} \mathcal{H}^{MN}\partial_N
    \mathcal{H}^{KL}\partial_L\mathcal{H}_{MK}
\end{align}
is called the generalized Ricci or curvature scalar and
\begin{equation}\label{genmetric}
  \mathcal{H}^{MN}=\begin{pmatrix}
    g_{ij} - B_{ik}g^{kl}B_{lj} & -B_{ik}g^{kj} \\
    g^{ik} B_{kj} & g^{ij}
  \end{pmatrix}
\end{equation}
is the generalized metric. It combines the metric $g_{ij}$ and the $B$-field $B_{ij}$ into an O($D,D$) valued, symmetric tensor fulfilling
\begin{equation}\label{eqn:genmetricO(D,D)}
  \mathcal{H}^{MN} \eta_{ML} \mathcal{H}^{LK} = \eta^{NK}\,.
\end{equation}
The dilaton $\phi$ of the NS/NS sector is encoded in the O(D,D) singlet
\be
d = \phi - \frac{1}{2}\log\sqrt{-g}
\label{generalizeddilaton}
\ee
It is called generalized dilaton.

\subsection{Generalized diffeomorphisms}
\label{generdiffeo}

The first symmetry of the action (\ref{dftaction}) is manifest global O(D,D) symmetry
\begin{equation}
  \mathcal{H}^{MN} \rightarrow \mathcal{H}^{LK} M_L{}^M M_K{}^N
    \quad \text{and} \quad
  X^M \rightarrow X^N M_N{}^M 
\end{equation}
where $M_L{}^K$ is a constant tensor which leaves $\eta^{MN}$ invariant, namely
\begin{equation}
  \eta^{LK} M_L{}^M M_K{}^N = \eta^{MN}\,.
\end{equation}
If this symmetry is broken to discrete O($D,D,\mathbb{Z}$) it can be interpreted as a T-duality transformation acting on the background torus where DFT is defined. This global symmetry can be extended to a local O(D,D) symmetry called generalized diffeomorphism. In infinitesimal form it is
\begin{equation}
  V^M \rightarrow V^M + \delta_\xi V^M
    \quad \text{and} \quad
  X^M \rightarrow X^M - \xi^M
\end{equation}
is mediated by the generalized Lie derivative
\begin{equation}\label{eqn:genliederiv}
  \delta_\xi V^M = \mathcal{L}_\xi V^M = \xi^N \partial_N V^M + (\partial^M \xi_N - \partial_N \xi^M) V^N
\end{equation}
and its generalization for higher rank tensors. The generalized metric $\mathcal{H}^{MN}$ and the generalized dilaton $d$ transform as
\begin{align}\label{eqn:genLieH}
  \delta_\xi \mathcal{H}^{MN} = \mathcal{L}_\xi \mathcal{H}^{MN} &=
    \xi^L \partial_L \mathcal{H}^{MN} + (\partial^M \xi_L - \partial_L \xi^M) \mathcal{H}^{LN}
    + (\partial^N \xi_L - \partial_L \xi^N) \mathcal{H}^{ML} \\
  \delta_\xi d = \mathcal{L}_\xi d &= \xi^M \partial_M d - \frac{1}{2} \partial_M \xi^M \,.
\end{align}
These transformations give rise to the algebra
\begin{equation}
  [\mathcal{L}_{\xi_1}, \mathcal{L}_{\xi_2}] = \mathcal{L}_{\xi_1} \mathcal{L}_{\xi_2} -
  \mathcal{L}_{\xi_2} \mathcal{L}_{\xi_1} = \mathcal{L}_{[\xi_1, \xi_2]_\mathrm{C}}
\end{equation}
which is governed by the C-bracket
\begin{equation}\label{eqn:Cbracket}
  \left[ \xi_1, \xi_2 \right]_\mathrm{C}^M = \xi_1^N \partial_N \xi_2^M -
    \frac{1}{2} \xi_{1 N} \partial^M \xi_2^N - \left( \xi_1 
    \leftrightarrow \xi_2 \right)\,,
\end{equation}
provided we impose the strong constraint
\begin{equation}\label{strongconstraint}
  \partial_N \partial^N \cdot = 0
\end{equation}
where $\cdot$ is a place holder for fields, gauge parameters and arbitrary products of them. In general, this algebra does not satisfy the Jacobi identity and so the generalized diffeomorphisms do not form a Lie algebroid. However, the failure to satisfy the Jacobi identity is a trivial gauge transformation which leaves all fields fulfilling the strong constraint invariant. A trivial way to solve \eqref{strongconstraint} is to set $\tilde \partial^i = 0$. In this case, the DFT action \eqref{dftaction} transforms into the low energy effective action
\begin{equation}
  S_\mathrm{NS} = \left. S_\mathrm{DFT} \right|_{\tilde \partial^i = 0} = \int\mathrm{d}^{D}x\,
    \sqrt{-g} e^{-2\phi} \big(\mathcal R + 4 \partial_\mu \phi \partial^\mu \phi 
    - \frac{1}{12} H_{\mu\nu\rho} H^{\mu\nu\rho} \big)
\end{equation}
There is a part in the effective action of all five superstring theories.

The DFT action (\ref{dftaction}) partial instead of covariant derivatives it is not manifestly invariant under generalized diffeomorphisms. But one can show that \cite{Hohm_H_Zw}
\begin{equation}\label{deltaxieR}
  \delta_\xi ( e^{-2d} \mathcal{R} ) = \partial_I ( \xi^I e^{-2d} \mathcal{R} )
\end{equation}
holds. Being a total derivative, this variation vanishes under the action integral and the action remains invariant. Like it is common in field theory, the variation $\delta_\xi$ commutes with partial derivative. It also is linear and obeys the Leibnitz rule.

\subsection{Double Lorentz symmetry}
\label{doublelorentzsymm}

There is a local double symmetry in DFT \cite{Siegel, Hohm_H_Zw, Hohm_K}. We write the generalized metric in terms of the generalized vielbein $E_A{}^M$ via
\begin{equation}\label{genmetricformvielbein}
  \mathcal{H}^{MN} = E_A{}^M S^{AB} E_B{}^N \,. 
\end{equation}
Using a frame formalism, we distinguish between flat and curved indices. The former are labeled by $A, B, \dots$ and the latter by $I, J, \dots\,$. 
In the flux formulation of DFT \cite{Geissbuhler, Aldazabal_M_N}, $E_A{}^M$ is an O($D,D$) valued matrix. 
Here, the flat generalized metric is given by
\begin{equation}
  S^{AB} = \begin{pmatrix} \eta_{ab} & 0 \\ 
      0 & \eta^{ab}
    \end{pmatrix}\,,
\end{equation}
where $\eta_{ab}$ and its inverse $\eta^{ab}$ denote the $D$-dimensional Minkowski metric. Due to the O($D,D$) valued generalized vielbein, the flat version
\begin{equation}\label{eqn:vielbeinodd}
  \eta^{AB} = E^A{}_M \eta^{MN} E^B{}_N
  \quad \text{with} \quad
  \eta^{AB}=\begin{pmatrix}
    0 & \delta_a^b \\
    \delta^a_b & 0
  \end{pmatrix}\,.
\end{equation}
of $\eta_{MN}$ does not differ from the curved one.

Now, consider the local double Lorentz transformation of the generalized vielbein
\begin{equation}\label{localodxodsym}
  E^A{}_M \rightarrow T^A{}_B E^B{}_M\,.
\end{equation}
We require that it leaves the generalized metric invariant. Hence, the transformation has to fulfill
\begin{equation}\label{trafoo2d}
  T^A{}_C S^{CD} T^B{}_D = S^{AB}\,.
\end{equation}
In addition, the transformed generalized vielbein has still to satisfy \eqref{eqn:vielbeinodd}, which gives rise to the further constraint
\begin{equation}\label{trafoodd}
  T^A{}_C \eta^{CD} T^B{}_D = \eta^{AB}\,.
\end{equation}
Transformations that simultaneously solve \eqref{trafoo2d} and \eqref{trafoodd} are of the form
\begin{equation}\label{trafoonxon}
  T^A{}_B = 
  \begin{pmatrix}
    u_a{}^b + v_a{}^b & u_{ab} - v_{ab} \\
    u^{ab} - v^{ab} & u^a{}_b + v^a{}_b
  \end{pmatrix}
\end{equation}
where $u_a{}^b$ and $v_a{}^b$ denote two independent O($1,D-1$) transformations with the defining properties
\begin{equation}
  u_a{}^c \eta_{cd} u_b{}^d = \eta_{ab}
  \quad \text{and} \quad
  v_a{}^c \eta_{cd} v_b{}^d = \eta_{ab}\,.
\end{equation}
By leaving the generalized and the $\eta$ metric invariant, double Lorentz transformations are a manifest symmetry of the action \eqref{dftaction}.

Except for the dilaton, the generalized vielbein combines all fields of the theory. As an element of O($D,D$) it has $D(2D-1)$ independent degrees of freedom. By gauge fixing the local double Lorentz symmetry only $D^2$ of them remain. A possible parameterization of the generalized vielbein is given by
\begin{equation}\label{eqn:EAMfixed}
  E^A{}_M = \begin{pmatrix}
    e_a{}^i & e_a{}^l B_{li} \\
    0 & e^a{}_i
  \end{pmatrix}
\end{equation}
in terms of the metric's vielbein $e^a{}_i$ with $e^a{}_i \eta_{ab} e^b{}_j = g_{ij}$ and the antisymmetric $B$-field $B_{ij}$. If $e^a{}_i$ is restricted to be an upper triangular matrix, this parameterization fixes the double Lorentz symmetry completely. An O(D,D) vielbein without any gauge fixing is
\begin{equation}\label{eqn:EAMgeneral}
  E^A{}_M = \begin{pmatrix}
    e_a{}^i & e_a{}^l B_{li} \\
    e^a{}_l \beta^{li} & e^a{}_i + e^a{}_l \beta^{lk} B_{ki}
  \end{pmatrix}
\end{equation}
where $e^a{}_i$ is an unrestricted vielbein of $g_{ij}$ and $\beta^{ij}$ is an antisymmetric bi-vector.

In the frame formalism of \cite{Siegel, Hohm_H_Z, Hohm_K}, the flat generalized metric reads
\begin{equation}
  S_{\bar A\bar B} = \begin{pmatrix}
    \eta_{a b} & 0 \\
    0 & \eta_{\bar a\bar b} \\
  \end{pmatrix}
\end{equation}
We have introduced two new set of indices: unbarred and barred indices. They are in one-to-one correspondence with the closed string's left- and right-moving parts. Two identical, D-dimensional Minkowski metrics $\eta_{ab}$, $\eta_{\bar a\bar b}$ and their inverse $\eta^{ab}$, $\eta^{\bar a\bar b}$ lower and raise them. The 2D-dimensional $\eta$ metric, which lowers and raises doubled indices, is defined as
\begin{equation}
  \eta_{\bar A\bar B} = \begin{pmatrix}
    \eta_{ab} & 0 \\
    0 & - \eta_{\bar a\bar b}
  \end{pmatrix}
  \quad \text{and} \quad
  \eta^{\bar A\bar B} = \begin{pmatrix}
    \eta^{ab} & 0 \\
    0 & - \eta^{\bar a\bar b}
  \end{pmatrix}\,.
\end{equation}
In order to link quantities in this frame formalism with the corresponding ones in the flux formulation, we apply the transformation
\begin{equation}\label{eqn:bartounbar}
  T^{\bar A}{}_B = \frac{1}{\sqrt{2}}
  \begin{pmatrix}
    u^{ba} & u_b{}^a \\
    -v^{b\bar a} & v_b{}^{\bar a} 
  \end{pmatrix} \quad \text{and} \quad
  T_{\bar A}{}^B = \frac{1}{\sqrt{2}}
  \begin{pmatrix}
    u_{ba} & u^b{}_a \\
    -v_{b\bar{a}} & v^b{}_{\bar a} 
  \end{pmatrix}\,,
\end{equation}
respectively. Again, $u_a{}^b$ and $v_a{}^{\bar b}$ are two independent Lorentz transformations fulfilling
\begin{equation}
  u_a{}^c \eta_{cd} u_b{}^d = \eta_{ab}
  \quad \text{and} \quad
  v_a{}^{\bar c} \eta_{\bar c\bar d} v_b{}^{\bar d} = \eta_{ab}\,.
\end{equation}
Applying \eqref{eqn:bartounbar} to the generalized and the $\eta$ metric, we obtain
\begin{equation}
  T^{\bar A}{}_C S_{\bar A\bar B} T^{\bar B}{}_D = S_{CD} 
    \quad \text{and} \quad
  T^{\bar A}{}_C \eta_{\bar A\bar B} T^{\bar B}{}_D = \eta_{CD} \,.
\end{equation}
Further, the generalized vielbein is
\begin{equation}
  E_{\bar A}{}^M = T_{\bar A}{}^B E_B{}^M = \frac{1}{\sqrt{2}} \begin{pmatrix}
    e_{ai} + e_a{}^l B_{li} & e_a{}^i \\
    - e_{\bar ai} + e_a{}^l B_{li} & e_{\bar a}{}^i \\
  \end{pmatrix}
\end{equation}
starting from \eqref{eqn:EAMfixed} and fixing $u_a{}^b = \delta_a^b$, $v_a{}^{\bar b}=\delta_a^{\bar b}$.

\section{The Extended Symmetry Model}
\label{extsymmod} 

Physically, the doubling of the local Lorentz symmetries indicates a genuine stringy character that there are two separate locally inertial frames, for both left and right moving modes \cite{Dimo_R_W, Hohm_H_Z}. For a point particle there is only one locally inertial frame. Without, introducing any additional physical degree of freedom, it is possible to reformulate this preon model as a DFT which couples to a gravitational background and manifests all the existing symmetries at once, including the O(4,4) T-duality, generalized diffeomorphism invariance, and a pair of local Lorentz symmetries, Spin(1,3)×Spin(3,1). These new symmetries are indicated by the action (\ref{dftaction}) and they are compatible with the previous Lagrangians (\ref{chirallagrangian}) and (\ref{chargedchiral})

The doubling of the local Lorentz symmetries implies that every fermion must belong to one of the two spin groups. In the present model there are two kinds of fermions: the preons (m) and the axino(n). Thus the obvious choice is that the preons belong to the Spin(1,3) and the axino to the Spin(3,1) representation. 

The model is consistent with the Big Bang model presented in \cite{Brand_C_F_W}. This model is singularity-free due to the near Hagedorn temperature thermal string condensate initial state. Thermal inflation from an string condensate fits in our particle scenario as well. 

\section{Conclusions}
\label{conclusions}

We have shown that, without any extra physical degree introduced, the present preon model can be readily reformulated as a Double Field Theory. The extended symmetry model couples to an arbitrary stringy gravitational background in an O(4,4) T-duality covariant manner and manifest two independent local Lorentz symmetries, Spin(1,3) × Spin(3,1). We are tempted to propose that this preon model is a pointlike low energy limit, or projection, of string theory. This model is a way to try to make connection between the SM particles and string theory, which has been an adversity in the top-down approach. 

The crucial test of the model is whether the SM superpartners exist or not. In the absence of SM superpartners the present model is a serious candidate for ontic particles. To humbly commemorate Steven Weinberg, who always had the big picture
in his mind, the motivation to the present note came from his text \cite{Weinberg_text}: ''Supersymmetry without superstrings is not an option''.

While finishing this note we found a paper discussing our string/particle connection hypothesis \cite{Lize_L_S_W}. The authors  formulate a new class of string amplitudes, and their corresponding spectra, by changing the boundary conditions of string propagators. The different boundary conditions are parametrized by a complex number $\lambda$ and are inspired by duality transformation of 4D Maxwell equations. The amplitudes and spectrum are of normal string theory when $\lambda = 1$. When $\lambda = -1$, the spectrum truncates to finite number and the amplitudes become a field theory (supergravity) amplitudes.

 


\vskip 1cm

\begin{thebibliography}{99}

\bibitem{Rai_00} Risto Raitio, A Model of Lepton and Quark Structure. Physica Scripta, 22, 197 (1980). \href{https://dx.doi.org/10.1088/0031-8949/22/3/002} {\tt PS22,197}. \href{http://vixra.org/abs/1903.0224} {viXra:1903.0224}
The model was conceived in November 1974 at SLAC. I proposed that the c-quark would be a gravitational excitation of the u-quark, both composites of three 'subquarks'. The idea was opposed by the community and was therefore not written down until five years later.

\bibitem{Rai_01} Risto Raitio, Supersymmetric preons and the standard model, Nuclear Physics B931 (2018) 283–290. \href{https://doi.org://10.1016/j.nuclphysb.2018.04.021} {\tt doi:10.1016/j.nuclphysb.2018.04.021} \href{https://arxiv.org/pdf/1805.03013.pdf} {\tt arXiv:1805.03013} 

\bibitem{Wess_Z} J. Wess and B. Zumino, Supergauge transformations in four dimensions, Nucl. Phys. B 70 (1974) 39. \href{https://doi.org/10.1016/0550-3213(74)90355-1} {\tt 10.1016/0550-3213(74)90355-1}

\bibitem{Figueroa-O} Jose Figueroa-O'Farrill, BUSSTEPP Lectures on Supersymmetry,
Lectures delivered at the 30th and 31st British Universities Summer Schools in Theoretical Elementary Particle Physics (BUSSTEPP) held in Oxford 2000 and in Manchester in 2001. \href{https://arxiv.org/pdf/hep-th/0109172.pdf} {\tt arXiv:hep-th/0109172}

\bibitem{Peccei_Q} Roberto D. Peccei and Helen R. Quinn,  CP Conservation in the Presence of Pseudoparticles, Phys. Rev. Lett. 38 (25) 1440–1443 (1977). 

\bibitem{Finkelstein} Robert Finkelstein, The SLq(2) extension of the standard model, Int. Journal of Modern Physics A, Vol. 30, No. 16, 1530037 (2015).

\bibitem{Marsh_00} David J. E. Marsh, Axions and ALPs: a very short introduction, 13th Patras Workshop on Axions, WIMPs and WISPs, Thessaloniki, May 2017. \href{https://arxiv.org/pdf/1712.03018.pdf} {\tt arXiv:1712.03018}

\bibitem{Marsh_01} David J. E. Marsh, Axion Cosmology, Physics Reports, Vol. 643 (2016). \href{https://arxiv.org/pdf/1510.07633.pdf} {\tt arXiv:1510.07633}

\bibitem{Baer_C_K_R} Howard Baer, Ki-Young Choi, Jihn E. Kim, Leszek Roszkowski, Dark matter production in the early Universe: beyond the thermal WIMP paradigm, Physics Reports 555 (2015) 1.

\bibitem{Marsh_et_al} Ren\'{e}e Hlozek, Daniel Grin, David J. E. Marsh, and Pedro G. Ferreira, A search for ultralight axions using precision cosmological data, Phys. Rev. D 91, 103512 (2015) \href{https://arxiv.org/pdf/1410.2896.pdf} {\tt arXiv:1410.2896}

\bibitem{DiBari} Pasquale Di Bari, On the origin of matter in the Universe. \href{https://arxiv.org/pdf/2107.13750.pdf} {\tt arXiv:2107.13750}

\bibitem{Cline} James M. Cline, Dark Atoms and Composite Dark Matter, Lectures given at Les Houches Summer School 2021: Dark Matter. \href{https://arxiv.org/pdf/2108.10314.pdf} {\tt arXiv:2108.10314}

\bibitem{Ketov_A_A} Yermek Aldabergenov, Andrea Addazi, and Sergei V. Ketov, Primordial black holes from modified supergravity. \href{https://arxiv.org/pdf/2006.16641.pdf} {\tt arXiv:2006.16641}

\bibitem{Fayet} Pierre Fayet, The U boson, interpolating between a generalized dark photon
or dark Z, an axial boson and an axionlike particle.  \href{https://arxiv.org/pdf/2010.04673.pdf} {\tt arXiv:2010.04673}

\bibitem{Wilczek} F. Wilczek, Problem of Strong P and T Invariance in the Presence of Instantons, Phys. Rev. Lett.. 40 (5): 279–282 (1978).

\bibitem{Weinberg_ax} Steven Weinberg, A New Light Boson?, Phys. Rev. Lett. 40 (4): 223–226 (1978). 

\bibitem{Guerra_M_P} Davide Guerra, Caio F. B. Macedo, and Paolo Pani, Axion boson stars, JCAP 09 (2019) 061. \href{https://arxiv.org/pdf/1909.05515.pdf} {\tt arXiv:1909.05515}

\bibitem{Annulli_C_V_0} Lorenzo Annulli, Vitor Cardoso, Rodrigo Vicente, Stirred and shaken: dynamical behavior of boson stars and dark matter cores. \href{https://arxiv.org/pdf/2007.03700.pdf} {\tt arXiv:2007.03700} 

\bibitem{Steinhardt_S} D.N. Spergel, P.J. Steinhardt, Phys. Rev. Lett. 84, 3760 (2000).

\bibitem{Steinhardt_W_D_F_M_S} Benjamin D. Wandelt, Romeel Dav\'{e}, Glennys R. Farrar, Patrick C. McGuire, David N. Spergel, and Paul J. Steinhardt, Self-Interacting Dark Matter, Proceedings of the International Conference DARK 2000 Heidelberg, Germany (2000). \href{https://arxiv.org/pdf/astro-ph/0006344.pdf} {\tt arXiv:astro-ph/0006344}

\bibitem{Hassler_L} Ralph Blumenhagen, Pascal du Bosque, Falk Hassler, Dieter Lust, Double Field Theory on Group Manifolds in a Nutshell,  Proceedings prepared for the Workshop on Geometry and Physics, Ringberg Castle, Germany (2016). \href{https://arxiv.org/pdf/1703.07347.pdf} {\tt arXiv:1703.07347}

\bibitem{Hassler_T} Falk Hassler, Double Field Theory on Group Manifolds (Thesis), LMU, München (2015). \href{https://arxiv.org/pdf/1509.07153.pdf} {\tt arXiv:1509.07153}

\bibitem{Hohm_H_Z} O. Hohm, C. Hull, and B. Zwiebach, Generalized metric formulation of double field theory, JHEP 1008 (2010) 008.\href{https://arxiv.org/pdf/1006.4823.pdf} {\tt arXiv:1006.4823}

\bibitem{Hohm_H_Zw} O. Hohm, C. Hull, and B. Zwiebach, Background independent action for double field theory, JHEP 1007 (2010) 016. \href{https://arxiv.org/pdf/1003.5027.pdf} {\tt arXiv:1003.5027}

\bibitem{Siegel} W. Siegel, Superspace duality in low-energy superstrings, Phys.Rev. D48 (1993) 2826–2837. \href{https://arxiv.org/pdf/hep-th/9305073.pdf} {\tt arXiv:hep-th/9305073}

\bibitem{Hohm_K} O. Hohm and S. K. Kwak, Frame-like Geometry of Double Field Theory, J.Phys. A44 (2011) 085404. \href{https://arxiv.org/pdf/1011.4101.pdf} {\tt arXiv:1011.4101}

\bibitem{Geissbuhler} D. Geissbuhler, Double Field Theory and N=4 Gauged Supergravity, JHEP 1111 (2011) 116. \href{https://arxiv.org/pdf/1109.4280.pdf} {\tt arXiv:1109.4280}

\bibitem{Aldazabal_M_N} G. Aldazabal, D. Marques, and C. Nunez, Double Field Theory: A Pedagogical Review, Class. Quant. Grav. 30 (2013) 163001. \href{https://arxiv.org/pdf/1305.190.pdf} {\tt arXiv:1305.1907}

\bibitem{Dimo_R_W} 10] S. Dimopoulos, S. Raby, and F. Wilczek, Supersymmetry and the Scale of Unification, Phys.Rev. D24 (1981) 1681–1683.

\bibitem{Brand_C_F_W} Robert Brandenberger, Renato Costa, Guilherme Franzmann, and Amanda Weltman,  T-dual cosmological solutions in double field theory, Phys. Rev. D 99, 023531 (2019). \href{https://arxiv.org/pdf/1809.03482.pdf} {\tt arXiv:1809.03482}

\bibitem{Lize_L_S_W} Matheus Loss Lize, Bochen Lyu, Warren Siegel, and Yu-Ping Wang, Chiral string theories as an interpolation between strings and particles. \href{https://arxiv.org/pdf/2109.10401.pdf} {\tt arXiv:2109.10401}

\bibitem{Weinberg_text} Steven Weinberg, The Quantum Theory of Fields, Volume 3: Supersymmetry. Cambridge University Press (2005).

\end{thebibliography}
\end{document}